\documentclass[prd,reprint,aps,tightenlines,preprintnumbers,amsmath,amssymb,showpacs,superscriptaddress]{revtex4-1}
\usepackage{amsmath,amssymb,graphicx,natbib}
\usepackage{graphicx}
\usepackage{dcolumn}
\usepackage{bm}
\usepackage[hyperindex,breaklinks]{hyperref}
\usepackage{amsmath}

\def \beq{\begin{equation}}
\def \eeq{\end{equation}}

\def\lsim{\mathrel{\rlap{\lower4pt\hbox{\hskip1pt$\sim$}}
    \raise1pt\hbox{$<$}}}                
\def\gsim{\mathrel{\rlap{\lower4pt\hbox{\hskip1pt$\sim$}}
    \raise1pt\hbox{$>$}}}                

\usepackage{relsize}
\def\babar{\mbox{\slshape B\kern-0.1em{\smaller A}\kern-0.1em
    B\kern-0.1em{\smaller A\kern-0.2em R}}}

\begin{document}

\title{Pion-Photon Transition Form Factor and New Physics in the $\tau$ Sector}

\author{David McKeen}
\email{mckeen@uvic.ca}
\affiliation{Department of Physics and Astronomy, University of Victoria, Victoria, BC V8P 5C2, Canada}

\author{Maxim Pospelov}
\email{mpospelov@perimeterinstitute.ca}
\affiliation{Department of Physics and Astronomy, University of Victoria, Victoria, BC V8P 5C2, Canada}
\affiliation{Perimeter Institute for Theoretical Physics, Waterloo, ON N2J 2W9, Canada}

\author{J. Michael Roney}
\email{mroney@uvic.ca}
\affiliation{Department of Physics and Astronomy, University of Victoria, Victoria, BC V8P 5C2, Canada}

\date{\today}

\begin{abstract}
Recent measurement of the $\gamma\gamma^\ast$ form factor of the neutral pion in the high $Q^2$ region disagrees with {\em a priori} predictions of QCD-based calculations.  We comment on existing explanations, and analyze a possibility that this discrepancy is not due to poorly understood QCD effects, but is a result of some new physics beyond the standard model (SM).  We show that such physics would necessarily involve a new neutral light state with mass close to the mass of $\pi^0$, and with stronger than $\pi^0$ couplings to heavier SM flavors such as $c$, $\tau$, and $b$.  It is found that only the coupling to the $\tau$ lepton can survive the existing constraints and lead to the observed rise of the pion form factor relative to $Q^{-2}$ at high $Q^2$.  We perform numerical fits to data and determine the allowed range of masses and couplings for such new particles.  This range of masses and couplings could also reduce or eliminate the tension between the $e^+e^-$ and $\tau$ decay determinations of the hadronic vacuum polarization.  Dedicated experimental analysis of $\tau$ pair production in association with such new states should provide a conclusive test of the new physics hypothesis as an explanation of the pion form factor rise.  We also comment on the calculations of the pion  form factor in the chiral quark model, and point out a possible dynamical origin of the quark mass scale inferred from the form factor measurement.
\end{abstract}

\maketitle


\begin{section}{Introduction}

The studies of properties of light pseudoscalar mesons via the ``photon fusion" 
mechanism has been a classic subject in particle physics 
that predates quantum chromodynamics (QCD). With the advent of QCD came the impetus 
to study the off-shell form factors of $\pi^0$, $\eta$, and $\eta^\prime$ as a way of testing QCD 
in exclusive processes. The most recent experimental advance in this direction has been 
achieved by the \babar\ collaboration \cite{Aubert:2009mc,delAmoSanchez:2011hk}
that measured the form factors of pseudoscalar mesons 
in the region of momentum transfer extending to $|Q^2| \lsim 100 ~{\rm GeV}^2$, significantly 
increasing the experimental $Q^2$ range over previous measurements by CLEO~\cite{Gronberg:1997fj} and CELLO~\cite{Behrend:1990sr}.  
The pion transition form factor measurement is based on
 the process $e^+e^-\to e^+e^-\gamma\gamma^\ast\to e^+e^-\pi^0$,
with one lepton scattered at a small angle and emitting a quasireal photon, 
and the other lepton scattering through a large angle, thus emitting a deeply
virtual $\gamma^*$.  No sign of convergence towards perturbative QCD 
asymptotics \cite{Lepage:1980fj} is seen in the \babar\ data for the $\pi^0$~\cite{Aubert:2009mc}, while the $\eta$ and $\eta^\prime$ form factors~\cite{delAmoSanchez:2011hk} 
appear to be consistent with prior theory expectations. 
The probed momentum transfers, $Q^2\sim 40$ GeV$^2$,  extend far beyond any nonperturbative 
energy scale relevant for low-energy QCD -- $Q^2 \gg m_\pi^2,~\Lambda_{\rm QCD}^2,~m_{\eta(\eta^\prime)}^2,~ (4\pi f_\pi)^2$, \dots -- which is why the disagreement of experimental results for $\pi^0$ with QCD-based theory expectations 
came as a total surprise.

An intense theoretical debate of the significance and interpretation 
of \babar\ result ensued, see, {\em e.g.}, Refs.~\cite{Dorokhov:2009jd,*Pham:2011zi,Radyushkin:2009zg,*Polyakov:2009je,*Wu:2010zc,*Agaev:2010aq,Arriola:2010aq,Dorokhov:2010bz,Roberts:2010rn,*Brodsky:2011yv,*Bakulev:2011rp}, 
and on the basis of this discussion the following logical possibilities can be identified: 

\begin{enumerate}

\item  The measurement of the $\pi^0$ form factor by \babar\ has some unidentified error.

\item Pre-\babar\ QCD predictions were somehow ``naive," and need to be augmented 
by assuming, {\em e.g.}, considerably different patterns of the quark momentum distribution within the $\pi^0$. 

\item   Alternative ways of describing the $\pi^0$ transition form factor, such as, {\em e.g.}, a 
constituent quark model, should be preferred over the QCD description.

\item  The discrepancy of the \babar\ measurement with QCD predictions 
is in fact a  signature of new physics beyond the standard model (SM).

\end{enumerate} 

While first three options on this list have been widely discussed in the literature, 
the last possibility, new physics, looks the least likely and thus far has been ignored. 
Indeed, a new physics option would have to be rather contrived: it must involve new 
particles/states below the weak scale, have interactions with photons and $\pi^0$ 
far stronger than the weak interactions, and evade all other constraints and tests. 

In this paper, we take this last option as an assumption and investigate 
broad classes of new physics capable of influencing the \babar\ form factor measurement. 
We identify two classes of models that lead to significant modifications of the 
form factor. In the first model, a pion mixing with a new light state results in the enhanced coupling 
of $\pi^0$ to intermediate mass quarks, $c$, $b$ and to the $\tau$ lepton. We call this model a ``hardcore"
pion. In the second class of models a new particle, scalar or pseudoscalar, is introduced with its mass very close to the $\pi^0$ mass. Such new particles can be called ``pion impostors." We show that the loop contributions of 
heavy quarks and $\tau$ leptons are capable of changing the form factor behavior, but 
only the coupling to $\tau$  has  a reasonable chance to remain undetected, while couplings to $b$ and $c$ 
quarks directly to $\pi^0$ or to a pion-like impostor particle are impossible to reconcile with bounds from 
$b\bar b$ and $c \bar c$ physics. 
We then identify  important search channels in $\tau$ pair production and decay 
which can be experimentally distinguished from SM backgrounds.
Therefore, these models can be decisively tested with the future analyses of the 
$\tau$ lepton dataset. 

This paper is organized as follows. Section~\ref{sec:QCD} introduces the basic formalism of form factor physics, discusses the benchmark QCD predictions, as well as alternative approaches within the chiral quark models. 
We choose the pre-\babar\ QCD prediction 
\cite{Bakulev:2001pa,*Bakulev:2001pb} that we use together with new physics 
contributions in our numerical fits. Section~\ref{sec:NP} introduces models of new physics and includes calculations of corrections 
to the form factor. In Sec.~\ref{sec:fits}, we determine the optimal ranges of parameters of the model, that is the strength of
couplings to $\tau$ leptons and the mass of the impostor particle. Section~\ref{sec:constraints} describes further constraints on the models and possible 
strategies in detecting a new (pseudo)scalar particle produced in association with $\tau$ pairs and in $\tau$ decays.  Our conclusions are reached in Sec.~\ref{sec:conclusions}. 
\end{section}


\begin{section}{Pion form factor in QCD and in quark models}
\label{sec:QCD}

We begin by introducing the matrix element for the relevant reaction, $\gamma^\ast\gamma^\ast\to P $, where $P$ stands for a pseudoscalar 
particle. It can be written as
\begin{align}
{\cal M}=ie^2{\cal F}\left(q^2,Q^2\right)\epsilon_{\mu\nu\rho\sigma}\epsilon_1^\mu\epsilon_2^\nu k_1^\rho k_2^\sigma
\end{align}
with $\epsilon_1$ and $\epsilon_2$ the polarizations of the photons, $k_1$ and $k_2$ their momenta, $k_1^2=-q^2$, $k_2^2=-Q^2$, and ${\cal F}\left(q^2,Q^2\right)$ the transition form factor. Calculations and measurements of ${\cal F}$ have a long history, 
and the form factor for the $\pi^0$-photon transition~\cite{Lepage:1980fj} at large photon virtualities 
is among the earliest exclusive calculations in perturbative QCD.    At large momentum transfers, the asymptotic freedom of QCD allows the matrix element to be factorized into the product of a hard scattering amplitude, $T$, calculated perturbatively, and a universal pion distribution amplitude (DA), $\phi_\pi$, which encodes nonperturbative effects:
\begin{align}
{\cal F}\left(q^2,Q^2\right)=\int_0^1dx~T\left(q^2,Q^2,\mu_F^2,x\right)\phi_\pi\left(x,\mu_F^2\right)
\end{align}
where $\mu_F$ is the factorization scale and $x$ is the momentum fraction carried by the quark in the pion. The DA is conveniently expanded in terms of Gegenbauer polynomials and obeys the Efremov-Radyushkin-Brodsky-Lepage (ERBL) evolution equation, with the asymptotic form $\phi_\pi^{\rm asy}\left(x,\mu_F^2\right)=\left(\sqrt{2}f_\pi/3\right)6x\left(1-x\right)$ 
where $f_\pi$ is the pion decay constant.   (We use the notation $f_\pi = \sqrt{2}F_\pi = 130.4$ MeV.)

Experimentally, it is easier to access the region where one of the photons in the reaction is nearly on-shell, $q^2\to0$, rather than both photons having large virtualities. Consequently, the large distance scale introduced by the on-shell photon necessitates use of nonperturbative techniques such as light-cone sum rules (LCSR) where ${\cal F}\left(q^2,Q^2\right)$ is related to a dispersion relation in $q^2$, resulting in an expression for the form factor ${\cal F}\left(Q^2\right)\equiv {\cal F}\left(q^2=0,Q^2\right)$ in terms of an integral over a spectral density.
Despite this complication, the pion form factor at sufficiently large $Q^2$ 
should approach the Brodsky-Lepage limit, ${\cal F}^{\rm BL} = \sqrt{2} f_\pi/Q^2 
\simeq 185~{\rm MeV}/Q^2$.
However, how exactly ${\cal F}$ reaches this limit as a function of $Q^2$ 
is open for debate, and many alternative ways 
of parameterizing the pion DA that could delay the onset of the 
asymptotic regime have been proposed over the years \cite{Chernyak:1981zz,*Chernyak:1981za,Bakulev:2001pa,*Bakulev:2001pb}.

A reasonable agreement of  CLEO~\cite{Gronberg:1997fj} and CELLO~\cite{Behrend:1990sr} results
with the QCD sum rule-derived pion form factor gave support to the
overall consistency of QCD-based theory and experiment.
A recent result from \babar\ on the $\eta$ and $\eta^{\prime}$ transition form factors is also in agreement
with QCD~\cite{delAmoSanchez:2011hk}.
 In light of this, the \babar\ result on the pion 
form factor increasing well above Brodsky-Lepage asymptotics and showing no sign of $Q^2{\cal F}(Q^2)$ 
``flattening," definitely looks puzzling. 
This increase in the form factor has spurred a large amount of theoretical work revisiting the issue.  
Notable post-\babar\ QCD-based explanations one way or another employ a ``flat" DA \cite{Radyushkin:2009zg,*Polyakov:2009je,*Wu:2010zc,*Agaev:2010aq}.
Also, serious doubts were voiced as to whether the \babar-observed growth of $Q^2F(Q^2)$ is actually 
possible within consistent theories of strong interactions \cite{Roberts:2010rn,*Brodsky:2011yv,*Bakulev:2011rp}. 
In our paper, we will not side with any of these claims, simply noting that none of QCD-based papers
could {\em predict} the pion form factor behavior, as observed by \babar\ above $Q^2 = 10~{\rm GeV}^2$. 
In light of this, in our subsequent new physics speculations, we are going to use the QCD form factor as predicted by 
the most elaborate {\em pre}-\babar\ QCD-based study in Ref.~\cite{Bakulev:2001pa,*Bakulev:2001pb}. 

The stark disagreement of the measured pion form factor with QCD  predictions has motivated some authors to 
evaluate the same quantity within various quark models. These models operate 
with ``elementary" pions and quarks, coexisting as dynamical degrees of freedom. Consider a quark model 
where the derivative coupling of $\pi^0$ to light quarks is governed  by $g_a= 1$ coupling, 
\begin{equation}
\label{cqm}
{\cal L}_{\rm QM} = \frac12\partial_\mu \pi^a \partial_\mu \pi^a + \bar q(iD_\mu \gamma_\mu - m_q) q
+\frac{\partial_\mu \pi^a}{2F_\pi} \bar q  \gamma_\mu \gamma_5 \tau^a  q ,
\end{equation} 
which is a subset of the full quark-meson Lagrangian \cite{Manohar:1983md,*Weinberg:2010bq}, and has explicit chiral symmetry in the limit $m_q \to 0$. 
Typically, these models assume that $m_q$ is given by the constituent quark mass scale, and therefore is not small. 
A simple calculation of a quark triangle diagram in the regime $Q^2 \gg m_q^2$ 
with two quark mass insertions gives well-known double-log asymptotics for the pion form factor, 
\begin{equation}
\label{naiveass}
{\cal F}_{\pi^0} (Q^2) = \frac{ 1}{4\pi^2F_\pi}\times \frac{ m_q^2}{Q^2} \log^2(Q^2/m_q^2)+\dots
\end{equation}
The current quark mass values of $O(10~{\rm MeV})$ provide a negligible contribution to the form factor at large $Q^2$,
but constituent quark masses on the order $O(100~ {\rm MeV})$ can provide a reasonable fit to the data, 
as was pointed out in a number of publications \cite{Dorokhov:2009jd,*Pham:2011zi,Arriola:2010aq}. 
The naive asymptotic formula of Eq.~(\ref{naiveass}) can be modified to account for the fact that 
beyond a certain nonperturbative scale $\Lambda$, the constituent mass $m_q$ is expected to get an additional 
power-like suppression, $m_q \sim \Lambda^2/Q^2$ \cite{Politzer:1976tv}, and that the pion-quark vertex may also 
be cutoff at certain quark virtualities. All this will lead to a stabilization of the logarithmic growth,
and in light of this the following phenomenological formula is preferred: 
\begin{equation}
{\cal F}_{\pi^0} (Q^2) = \frac{ 1}{4\pi^2F_\pi}\times \frac{ m_q^2}{Q^2} \left\{ \begin{array}{c}
\log^2(Q^2/m_q^2)~~{\rm for}~Q^2\lsim \Lambda^2\\
\log^2(\Lambda^2/m_q^2) ~~{\rm for}~Q^2\gsim \Lambda^2
\end{array}
\right..
\label{lessnaive}
\end{equation}
The functional form of the transition between these two regimes is of course impossible to track without 
extra assumptions about the $m_q(Q^2)$ behavior. 

A fit to the experimental data can provide information on the preferred values for the parameters $m_q$ and $\Lambda$.  To interpolate between the two regions of Eq.~\ref{lessnaive}, we take $Q^2\Lambda^2/m_q^2(Q^2+\Lambda^2)$ as the argument of the logarithm.  
We plot the results of such a fit in Fig.~\ref{fig:varmassfit} where 3$\sigma$ contours of preferred regions 
are shown when all experimental pion form factor data are used (black, solid), and when only the \babar\ 
results are employed (red, dashed). This figure shows that when all data are included in the fit, the preference is for 
$m_q$ in the range from 130 to 200 MeV, and $\Lambda$ between 2 and 5 GeV.  When only \babar\ 
data are included, there is a clear preference for larger $\Lambda $, including values approaching infinity, 
in which case the quark masses should be in a relatively narrow range around  $m_q\simeq 130$ MeV. This is 
consistent with results found by other authors \cite{Dorokhov:2009jd,*Pham:2011zi}. 

\begin{figure}
\rotatebox{0}{\resizebox{70mm}{!}{\includegraphics{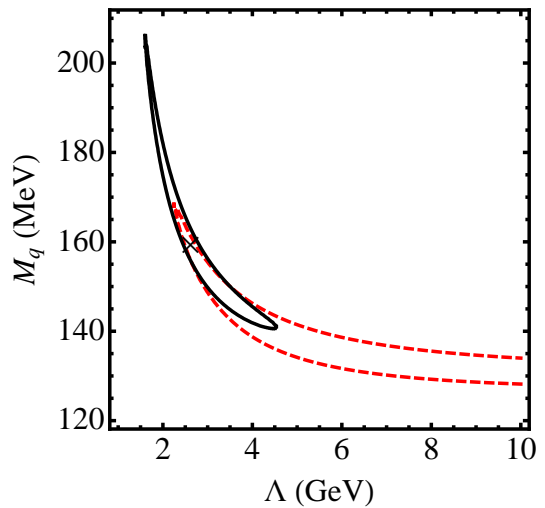}}}
\caption{3$\sigma$ contours resulting from the fit of Eq.~(\ref{lessnaive})
to the experimental extraction of pion form factor. The solid (black) contour corresponds to using
all experimental data, and the dashed (red) to using only \babar\ data. The best-fit point marked inside the black 
contour is at $m_q \simeq 160$ MeV and $\Lambda \simeq 2.6$ GeV.}
\label{fig:varmassfit}
\end{figure}

How legitimate are such explanations? Some authors \cite{Dorokhov:2010bz}
remark that a mass scale of 130 MeV looks very odd: it is far above the current 
quark mass values, and yet significantly below a ``typical" constituent mass value of $\sim$300 MeV. 
Moreover, if both light quarks and pions are included within the same dynamical scheme, one should 
be worried about the self-consistency of the theory at the loop level, renormalization of $F_\pi$, etc. 

Here we would like to point out how a mass scale $m_q \sim 130$ MeV 
could emerge in the theory described by (\ref{cqm}). We consider the leading order 
power correction to the propagator of a light quark in QCD and in the chiral quark model. 
For simplicity, we take the value of the 
current quark mass $\overline{m} = (m_u+m_d)/2$ 
close to zero ($m_\pi\to 0$ limit), while maintaining a 
finite value of the quark vacuum condensate $\langle \bar qq \rangle\sim - (270 ~{\rm MeV})^3$. 
For the case of two light flavors with SU(2) symmetry, 
direct calculation show that the quark propagators can be written as
\begin{eqnarray} 
\label{QCD}
S_{\rm QCD}(p) = \frac{i}{p\!\!\!/} + \frac{i 4\pi\alpha_s \langle \bar qq \rangle }{3p^4}+\dots,\\
S_{\rm CQM}(p) = \frac{i}{p\!\!\!/} - \frac{i  \langle \bar qq \rangle }{16F_\pi^2p^2}+\dots
\label{us}
\end{eqnarray}
where ellipses stand for higher-order power corrections in terms of the ratio of nonperturbative 
scale of strong interaction to $p^2$. 
The power corrections can be matched to the $S(p) = i/p\!\!\!/ +im_q^{\rm eff}/p^2$, and the second terms in these 
formulae can be identified with the quark mass contribution. 
The QCD result (\ref{QCD}) is of course a well-known formula due to Politzer in the Landau gauge
\cite{Politzer:1976tv,Pascual:1981jr} (with the numerical error of 
\cite{Politzer:1976tv} corrected), resulting from the 
quark-gluon propagator mixing, Fig.~\ref{fig:condensate}.
The second line, Eq.~(\ref{us}), is the result of the quark-meson propagator 
mixing in the chiral quark model. 
The scale of the effective quark mass suggested by (\ref{us}) is
\begin{equation}
\label{mq}
m_q^{\rm eff} = -\frac{  \langle \bar qq \rangle }{16F_\pi^2} = \frac{m_\pi^2}{16(m_u+m_d)} \simeq 135 ~{\rm MeV }\times 
\frac{4.3 ~{\rm MeV}}{\overline{m}}
\end{equation}
where the current quark mass value is within the range of $3.0-4.8$ MeV at a normalization point of $2$ GeV, 
as quoted by the Particle Data Group \cite{Nakamura:2010zzi}. Interestingly enough, the mass scale emerging from the 
quark condensate  correction to the quark propagator (\ref{mq}) is
 exactly in the range suggested by our fit to the pion form factor. If instead one considers 
three flavors of light quarks in the chiral limit with flavor SU(3) symmetry, the result changes to $
\langle\bar qq \rangle/(9F_\pi^2) \simeq 240$ MeV. 
We stress that (\ref{mq}) is an effective quark mass that neither breaks chiral symmetry explicitly 
nor makes the pion massive.

The QCD result is clearly inapplicable below 1 GeV, and moreover, it 
has an explicit gauge dependence, that will cancel in 
the full one-loop evaluation of the gauge-independent correlators of quark currents \cite{Pascual:1981jr}.  
The result for $S_{\rm CQM}$ (\ref{us}) is of course gauge independent. Because of this distinction, 
the matching of $S_{\rm CQM}$ and $S_{\rm QCD}$ cannot be done rigorously.
Nevertheless, the power corrections in (\ref{QCD}) and (\ref{us}) become comparable around 
$p^2 \sim 1~{\rm GeV}^2$, which is precisely the scale where one would expect the 
quark model description to become invalid and be replaced by perturbative QCD. 
One can see that this scale is a lot lower than scale $\Lambda$ 
suggested by the numerical fits to the form factor. 

The closeness of $m_q^{\rm eff}$ in Eq.~(\ref{mq}) to the value suggested by the chiral quark model interpretations 
of the pion form factor is intriguing. However, at this point it cannot be considered as a legitimate explanation
of the \babar-observed rise of the pion form factor. One has to use the chiral quark model well beyond $|p^2|\sim 1$ GeV$^2$ 
and indeed outside of its usual range of validity. Moreover, even assuming that quark-pion Lagrangian can be extended that far
in energy, one would still need to calculate the chiral loop corrections to the triangle diagram to see at what energy scale the
whole framework becomes  inconsistent. Within SU(3) flavor symmetry, it is also difficult, if not impossible, to explain a 
markedly different behavior of the form factors for the pion and the SU(3) octet combination of $\eta$ and $\eta^\prime$~\cite{Wu:2011gf,*Lucha:2011if,*Balakireva:2011wp}.
Nevertheless, the emergence of the effective quark mass due to the quark condensate, Eq.~(\ref{mq}), 
within massless chiral quark models calls for the evaluation of various quark current correlators in the same framework. 
A procedure very similar to the QCD operator product expansion (OPE) can be developed this way, with the hope that 
(\ref{cqm}) could provide a reasonable description of hadronic correlators in the intermediate range of energies, 
$(m_q^{\rm eff})^2 \ll |p^2| \ll {\rm GeV}^2$. We will leave the investigation of the hadronic models with pions on the 
``wrong side" of the OPE to future work. 

To summarize this section, QCD calculations of the pion form factor were remarkably unsuccessful in 
predicting the rise of the pion form factor observed by \babar. Chiral quark models may provide a reasonable explanation, but have to be 
extended far beyond the usual limits of their validity. 

\begin{figure}
\rotatebox{0}{\resizebox{70mm}{!}{\includegraphics{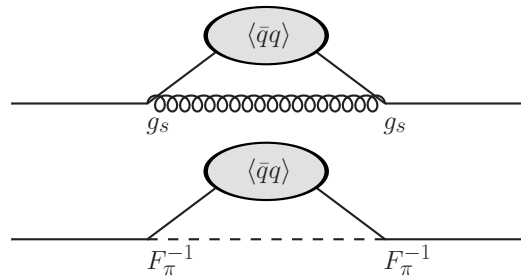}}}
\caption{Nonperturbative corrections to the quark propagator in QCD and in the chiral 
quark models. }
\label{fig:condensate}
\end{figure}

\end{section}


\begin{section}{New Physics}
\label{sec:NP}
In this section, we describe an explanation of the rise in the $\pi^0$-$\gamma$ transition form factor using new physics.  
We will consider $\tau$ leptons coupled to either new pseudoscalar or scalar degrees of freedom with masses close to that of the neutral pion.  
In the case of a new pseudoscalar, there is also the possibility of mixing with the $\pi^0$ so that an effective coupling of $\pi^0$ to $\tau$ is induced.
As will become clear later, new neutral particles coupled to $\tau$-leptons is perhaps the only model of new physics that can provide an 
appreciable contribution to the pion transition form factor and evade other constraints.

The masses of the new light states we consider are essentially free parameters in this phenomenological study and only fixed according to experimental constraints.  Their small masses do not necessarily result from their being psuedo-Nambu-Goldstone bosons of a spontaneously broken global symmetry as in the case of the pions.  A precise embedding of the models we consider in a more complete theory could, however, offer an explanation.


\begin{subsection}{A New Pseudoscalar}
We first consider a SM singlet pseudoscalar, $P$, coupled to SM fermions through dimension-5 operators,
\begin{align}
{\cal L}_{\rm int}&=-iP\left(\frac{H^0}{\Lambda}\right)\left(\lambda_\tau \bar \tau \gamma^5\tau+\lambda_u \bar u \gamma^5u+\lambda_d \bar d \gamma^5d+\dots\right),
\label{eq:PInt}
\end{align}
where $H^0$ is the neutral component of the SM Higgs doublet and $\Lambda$ is the scale at which this effective interaction is generated. 
The ellipsis refers to other couplings, {\em e.g.}, to other SM fermions, which we will ignore for the time being.  Equation~(\ref{eq:PInt}) is the unitary gauge expression for the 
${\rm SU}_{\rm L}(2)\times {\rm U}_{\rm Y}(1)$-invariant effective interaction.  When the Higgs field is replaced by its vacuum expectation value $v/\sqrt{2}$, effective Yukawa interactions result,
\begin{align}
{\cal L}_{\rm int}&=-iP\left(g_\tau \bar \tau \gamma^5\tau+g_u \bar u \gamma^5u+g_d \bar d \gamma^5d\right),
\label{eq:PeffInt}
\end{align}
with $g_u=\lambda_u v/\sqrt{2}\Lambda$, etc.  Since we expect $\Lambda$ to be near the scale of electroweak symmetry breaking, the effective Yukawa couplings $g_{u,d,\tau}$ might not be small unless required by experiment. 

An axial-vector coupling and a photon coupling are available for the pseudoscalar particles as well: $f_{\tau}^{-1}\partial_\mu P \bar\tau \gamma_\mu\gamma_5 \tau$ and $f_{\gamma}^{-1}PF_{\mu\nu}\tilde F_{\mu\nu}$. For each flavor of fermions these three couplings are 
related by one anomaly relation. We choose to eliminate the axial-vector couplings and keep the pseudoscalar and photon couplings. 
Moreover, the ``contact" photon coupling is not suitable for fitting the pion form factor, as it gives a $Q^2$-independent contribution 
to ${\cal F}(Q^2)$ to be compared with the fiducial $1/Q^2$ fall off. 

We can now imagine two scenarios distinguished by whether there is mixing between $P$ and $\pi^0$.


\begin{subsubsection}{Pseudoscalar Impostor}
First, if the Yukawa couplings of $P$ to $u$ and $d$ are vanishing, mixing between $P$ and $\pi^0$ is negligible, and the only interaction involving $P$ and SM fields that is important is
\begin{align}
{\cal L}_{\rm int}&=-ig_\tau P \bar \tau \gamma^5\tau.
\label{eq:lag_P}
\end{align}
This leads to the following amplitude for $\gamma\gamma^\ast\to P$ through a $\tau$ loop,
\begin{align}
{\cal M}_{P\gamma\gamma^\ast}=&ie^2{\cal F}_P\left(Q^2\right)\epsilon_{\mu\nu\rho\sigma}\epsilon_1^\mu\epsilon_2^\nu k_1^\rho k_2^\sigma,
\label{eq:MgamgamP}
\end{align}
where $k_1$ and $k_2$ are the photons' momenta with $k_1^2=-q^2=0$ and $k_2^2=-Q^2$.  The form factor here is given by
a well-known result for the fermion triangle diagram \cite{Ametller:1983ec}
\begin{align}
&{\cal F}_P\left(Q^2\right)=\frac{g_{\tau}m_\tau}{\pi^2\left(Q^2+m_P^2\right)}\left[L^2+\left(\sin^{-1}\frac{m_P}{2m_\tau}\right)^2\right],
\label{FFP}
\end{align}
with
\begin{align}
L=\frac{1}{2}\log\left(\frac{R_{Q^2}-1}{R_{Q^2}+1}\right),~~~R_{Q^2}=\sqrt{1+\frac{4m_\tau^2}{Q^2}}.
\label{eq:Ldef}
\end{align}
If kinematically allowed, $P$ would be produced in $e^+e^-\to e^+e^-\gamma\gamma^\ast\to e^+e^-P$ and primarily decay to $\gamma\gamma$.  Further, if the mass of the pseudoscalar is tuned so that $m_{P}\simeq m_{\pi^0}$ (a quantitative study of the extent to which this must hold will be given in Sec.~\ref{sec:mass}) then $P$ would be misidentified as a $\pi^0$ when reconstructed using a peak in the $M_{\gamma\gamma}$ spectrum.  The $\pi^0$-$\gamma$ form factor is extracted from the measurement of the cross section for $e^+e^-\to e^+e^-\pi^0$ using
\begin{align}
\frac{d\sigma\left(e^+e^-\to e^+e^-\pi^0\right)}{dQ^2}= \pi\alpha^4 \frac{\left|{\cal F}\left(Q^2\right)\right|^2}{Q^2} A\left(Q^2\right)
\end{align}
where $A\left(Q^2\right)$ is a function that depends on the particulars of the experiment and on radiative corrections.  The matrix element in Eq.~(\ref{eq:MgamgamP}) leads to a cross section for $e^+e^-\to e^+e^-P$ given by $ \pi\alpha^4\left(\left|{\cal F}_P\left(Q^2\right)\right|^2/Q^2\right)A\left(Q^2\right)$ which, since $m_{P}\simeq m_{\pi^0}$, is interpreted as an additional contribution to the $\pi^0$ production cross section.  This would lead to a modification of the form factor extracted,
\begin{align}
\left|{\cal F}\left(Q^2\right)\right|\to\sqrt{\left|{\cal F}\left(Q^2\right)\right|^2+\left|{\cal F}_{P}\left(Q^2\right)\right|^2}.
\label{eq:PImPP_FF}
\end{align}
We will refer to $P$ in this scenario as a ``pseudoscalar impostor." In this formula, we also assume that the contact photon coupling 
of $P$ to $F_{\mu\nu}\tilde F_{\mu\nu}$ is small. If it is not, then the ${\cal F}_P(Q^2)$ contribution in (\ref{eq:PImPP_FF}) will receive a 
constant $Q^2$-independent addition. 
\end{subsubsection}


\begin{subsubsection}{A Hardcore Pion}
The second scenario that could result from the interaction in Eq.~(\ref{eq:PeffInt}) is if the Yukawa couplings to $u$ and $d$ quarks are  large enough to lead to appreciable mixing between $P$ and $\pi^0$ if, again, $m_{P}\sim m_{\pi^0}$.

In this case, an effective interaction of the $\pi^0$ itself with the $\tau$ is generated,
\begin{align}
{\cal L}_{\rm int}=-ig_\tau^{\pi}\pi^0\bar\tau\gamma^5\tau,
\label{eq:lag_pi}
\end{align}
where $g_\tau^\pi=g_\tau\sin\theta\simeq g_\tau\theta$, and the effective mixing angle between $P$ and $\pi^0$ is given by
\begin{equation}
\theta =\frac{|\langle\bar qq\rangle|(g_u-g_d) }{F_\pi(m_{\pi^0}^2-m_P^2)} \simeq 
\frac{\left(400~{\rm MeV}\right)^2\times (g_u-g_d) }{2m_{\pi^0}\Delta m_{P\pi^0}} .
\end{equation} 
We shall take the mass splitting to be small, $\Delta m_{P\pi^0}=\left|m_{\pi^0}-m_P\right|\ll m_\pi$.

As in the no-mixing case above, besides a correction to $\gamma\gamma^*\to \pi^0$, there will also be an amplitude for $\gamma\gamma^\ast\to P$, potentially complicating the analysis.  To keep things simple, in this case with mixing between $P$ and $\pi^0$, we can make the production of real $P$'s experimentally unimportant by adding a long lived singlet state, $\phi$ coupled to $P$, so that $P$ decays invisibly, $P\to\phi\phi$.  To avoid making the invisible width of the $\pi^0$ too large, we can forbid the decay $\pi^0\to\phi\phi$ kinematically if, {\em e.g.}, the mass of $\phi$ is $70~{\rm MeV}$ and $m_P=145~{\rm MeV}$.  This way, the only change in the apparent form factor in the mixing case is given by Eq.~(\ref{eq:HCP_FF}) and we do not deal with the added complication of a second state in the $\gamma\gamma$ spectrum just as we do in the no-mixing (pseudoscalar impostor) case. 

Given the interaction in Eq.~(\ref{eq:lag_pi}), there will be a change in the actual $\pi^0$ form factor.  This change is simply
\begin{align}
{\cal F}\left(Q^2\right)\to {\cal F}\left(Q^2\right)+{\cal F}_{P}\left(Q^2\right),
\label{eq:HCP_FF}
\end{align}
where ${\cal F}_{P}\left(Q^2\right)$ is given by Eq.~(\ref{FFP}) with the replacements $g_{\tau}\to g_\tau^{\pi}$, $m_{P}\to m_{\pi^0}$.  We will call this the ``hardcore pion" scenario.

Because the hardcore pion model uses couplings of new states 
to light quarks it should be considered ``more extreme" than the pseudoscalar impostor model, and is subject to additional constraints coming from the quark sector.  Therefore, it deserves special discussion about its viability. The angle $\theta$ characterizes the admixture of the new $P$ state in the physical $\pi^0$. As such, the well-tested rate for the decay
 $\pi^\pm\to\pi^0 e^\pm \nu$ will be corrected,
\begin{equation}
\Gamma\left(\pi^\pm\to\pi^0 e^\pm \nu\right) = \Gamma_{\rm SM}\times \cos^2\theta \simeq
\Gamma_{\rm SM}\times (1-\theta^2/2).
\end{equation}
From the 1\% agreement between theory and experiment for this rate one should conclude that $\theta^2<0.02$. To satisfy this limit
with $m_P= 145$ MeV, the Yukawa couplings should be at the level of $|g_u-g_d| < 2\times 10^{-3}$, which is not a 
very tight bound. The mass shift of the physical pion is $\sim \Delta m_{P\pi^0} \theta^2$, and is less than 0.2 MeV if
$\Delta m_{P\pi^0} \leq 10$ MeV, within the uncertainty of the theoretical calculation of the neutral-charged pion mass difference (see, {\em e.g.}, \cite{Portelli:2010yn}).

Since we do not impose any symmetry requirements on $g_{u,d}$,  then one should also expect additional violations of isospin invariance due interactions mediated by $P$. The measure of such new effects is on the order of 
$g^2_{u,d}/(m_q/F_\pi)^2$, which again should not exceed $\sim 1\%$.  One can  see that with $g_{u,d}\sim O(10^{-3})$
such constraints will be easily satisfied. The effects of a possible mixing with $\eta$ are not 
important, as we assume that $P$ is much closer in mass to $\pi^0$.

Another important constraint could come from missing energy decays of 
$K$ mesons, $K^+\to \pi^+ P\to \pi^+ E \!\!\!\!/$, since we assume that $P$ decays invisibly.  The branching ratio for this decay is tested to $O(10^{-10})$~\cite{Artamonov:2008qb}, and indeed could provide a powerful constraint, except in the region of 
parameter space with $|m_{\pi^0}-m_P|\leq 10$ MeV, where the measurement of the missing energy rate is not possible 
due to $\pi^+\pi^0$ background.

We conclude that based on the arguments of isospin invariance, precision tests of 
charged pion beta decay, and the missing energy decay of $K^+$, the mixing angle $\theta$ is limited to $\sim 0.15$, 
while the mass of a new pseudoscalar state should be close to the pion mass within $\sim 10$ MeV. 
Finally, the effect of this mixing on the rate for $\pi^0\to\gamma\gamma$ is discussed in Sec.~\ref{sec:pi0width}.


\end{subsubsection}

\end{subsection}


\begin{subsection}{A New Scalar}
If we consider a new scalar instead of a pseudoscalar, the analysis is similar but simplified.  Since the scalar mixing with $\pi^0$
breaks CP symmetry, we ignore couplings to light quarks and only consider couplings of the scalar to $\tau$ through
\begin{align}
{\cal L}^S_{\rm int}=-S\left(\frac{H^0}{\Lambda}\right)\left(\lambda^\prime_\tau \bar \tau \tau\right)\to-h_{\tau}S\bar\tau\tau.
\label{eq:lag_S}
\end{align}
As in the pseudoscalar case, this generates an amplitude for two photons to produce a scalar, $\gamma\gamma^\ast\to S$,
\begin{align}
{\cal M}_{S\gamma\gamma}=&e^2{\cal F}_S\left(Q^2\right)\left(k_1^\nu k_2^\mu- k_1\cdot k_2 g^{\mu\nu}\right)\epsilon_{1\mu}\epsilon_{2\nu}
\label{eq:MgamgamS}
\end{align}
and we have again taken $k_1^2=-q^2=0$.  The form factor in this case is
\begin{align}
&{\cal F}_S\left(Q^2\right)=\frac{h_{\tau}m_\tau}{\pi^2\left(Q^2+m_S^2\right)}
\nonumber
\\
&\times\Bigg\{\left(1-\frac{4m_\tau^2}{Q^2+m_S^2}\right)\left[L^2+\left(\sin^{-1}\frac{m_S}{2m_\tau}\right)^2\right]
\label{FFS}
\\
&~~+\frac{2Q^2}{Q^2+m_S^2}\left[R_{Q^2}L+R_{m_S^2}\left(\sin^{-1}\frac{m_S}{2m_\tau}\right)\right]+1\Bigg\}
\nonumber
\end{align}
with $L$ and $R_{Q^2}$ as in Eq.~(\ref{eq:Ldef}) and
\begin{align}
R_{m_S^2}=\sqrt{\frac{4m_\tau^2}{m_S^2}-1}.
\end{align}
Defined in this way, the $e^+e^-\to e^+e^-S$ cross section is
\begin{align}
\frac{d\sigma\left(e^+e^-\to e^+e^-S\right)}{dQ^2}= \pi\alpha^4 \frac{\left|{\cal F}_S\left(Q^2\right)\right|^2}{Q^2} A\left(Q^2\right)
\end{align}
with $A\left(Q^2\right)$ the same as in the pseudoscalar case.  Therefore, if the scalar mass is close to that of the neutral pion, it could mimic a $\pi^0$ giving an apparent contribution to the $\pi^0$-$\gamma$ form factor,
\begin{align}
\left|{\cal F}\left(Q^2\right)\right|\to\sqrt{\left|{\cal F}\left(Q^2\right)\right|^2+\left|{\cal F}_{S}\left(Q^2\right)\right|^2}.
\label{eq:PImPS_FF}
\end{align}
This particle will be referred to as a ``scalar impostor."
\end{subsection}


\begin{subsection}{Why $\tau$?}
\label{sec:tau}
We now turn our attention to the motivation for choosing $\tau$ in the interactions in Eqs.~(\ref{eq:lag_P}), (\ref{eq:lag_pi}), and (\ref{eq:lag_S}).  First, if we wish to have a change in the (apparent) $\pi^0$-$\gamma$ form factor due to a loop-induced coupling of an impostor or the pion to $\gamma\gamma$, the new states running in the loop must obviously be electrically charged.  If it is a state unaccounted for in the SM, then its mass must be at least on the order of $100~{\rm GeV}$ to escape collider bounds on pair production of new charged particles.  Particles with masses this large will require extremely large couplings to $P$ or $S$ to have any effect on $F\left(Q^2\right)$ for $Q^2\sim 10-40~{\rm GeV}^2$. Thus, we are led to only consider charged SM fermions (excluding the top quark) as candidates for coupling to impostors or to the $\pi^0$.

In order to change the form factor behavior, the couplings of new states to light quarks would have to be very large, stronger than the pion coupling 
to quarks, which is clearly unrealistic. Therefore, we ignore couplings to $u,d,s$, only considering small values that potentially mediate mixing with the pion in the hardcore pion case but are otherwise negligible.
 
Couplings to charm and bottom quarks could explain the rise in the pion form factor but run into limits from the decays of $c\bar c$ and $b\bar b$ states.

In the charm quark case, the couplings can be constrained by data on $J/\psi\to\gamma\pi^0$ and $\psi^\prime\to\gamma\pi^0$ since pion impostors that decay to $\gamma\gamma$ will be reconstructed as neutral pions.  Requiring that ${\cal B}\left(\psi^\prime\to\gamma\pi^0\right)$not exceed its experimental value of $\left(1.58\pm0.42\right)\times10^{-6}$ ~\cite{Ablikim:2010dx} sets limits on the coupling of impostors or a hardcore pion to $c$ quarks of
\begin{align}
\left|g_{c}\right|,\left|g_c^{\pi}\right|\lsim 0.018,~~~\left|h_{c}\right|\lsim 0.023
\end{align}
which are too small to have an appreciable effect on the $\pi^0$ form factor.

To constrain the coupling to bottom quarks, we note that $b$ quarks interacting directly with the $\pi^0$ or a scalar or pseudoscalar impostor will lead to (apparent) enhancements in the transitions between $b\bar b$ states with the emission of a $\pi^0$.  Approximating the interquark potential as being Coulombic and using the experimental limit, ${\cal B}\left(\Upsilon(2S)\to\Upsilon(1S)\pi^0\right)<1.8\times10^{-4}$ at 90\% C. L.~\cite{He:2008xk}, results in the limits on the couplings to $b$ quarks of
\begin{align}
\left|g_{b}\right|,\left|g_b^{\pi}\right|\lsim 0.055,~~~\left|h_{b}\right|\lsim 0.0055.
\end{align}
As in the case of the charm quark, these are too small to give a non-negligible contribution to the form factor.

Moving on to leptons, we can immediately rule out the electron as a candidate since this will lead to $\gamma\gamma$ no longer being the dominant decay mode for $P$ or $S$ in the impostor cases, and in the hardcore pion case leading to a branching ratio for $\pi^0\to e^+ e^-$ that is many orders of magnitude larger than experimentally allowed.

The strictest limits on couplings to muons come from  $(g-2)_\mu$.  These couplings must be suppressed to the level of $10^{-3}$ to $10^{-4}$, and therefore coupling to the muon cannot offer an explanation of the rise in the form factor.

For these reasons, we concentrate on the scenario where the $\pi^0$ or an impostor is coupled to $\tau$.
\end{subsection}


\begin{subsection}{Limits from $\left(g-2\right)_\tau$}
\label{sec:g-2}
The most immediate constraints on the interactions of the $\tau$ with light spin-0 particles, as in the case of the muon, are due to limits on the its anomalous magnetic moment but, since it is much less precisely known than that of the muon, the constraints are weaker.  Pseudoscalars or scalars whose masses are equal to $m_{\pi^0}$ interacting with $\tau$ as in Eq.~(\ref{eq:lag_P}) or (\ref{eq:lag_S}) will induce contributions to $a_\tau=(g-2)_\tau/2$ of~\cite{Leveille:1977rc,*McKeen:2009ny}
\begin{align}
\left(\Delta a_\tau\right)_P&=-6.21\times 10^{-3} g_\tau^2,
\\
\left(\Delta a_\tau\right)_S&=1.66\times 10^{-2} h_\tau^2.
\end{align}
The hardcore pion scenario results in the same contribution as in $\left(\Delta a_\tau\right)_P$ with $g_\tau\to g_\tau^\pi$.
DELPHI provides the most stringent experimental limits on $a_\tau$: 
$-0.052 < a_\tau < 0.013$ at 95\% confidence level~\cite{Abdallah:2003xd}.
  Taking the SM expectation into account and requiring that the total $a_\tau$ not fall outside of this range gives the following limits on the couplings:
\begin{align}
\left|g_\tau\right|,\left|g_\tau^\pi\right|\lsim 2.9;~~~\left|h_\tau\right|\lsim 0.84.
\end{align}
These limits are mild enough that coupling scalars or pseudoscalars to $\tau$ can offer an explanation for the rise in the $\pi^0$-$\gamma$ form factor above the asymptotic value in QCD.  
We will postpone discussion of other constraints on coupling spin-0 particles to $\tau$ such as higher-order corrections to $(g-2)_{e,\mu}$, limits from $Z\to\gamma\pi^0$ and $Z\to\gamma\tau^+\tau^-$, and $\tau$ decays until Sec.~\ref{sec:constraints}.
\end{subsection}
\end{section}


\begin{section}{Fitting the $\pi^0$-$\gamma$ Form Factor Data}
\label{sec:fits}
\begin{subsection}{Couplings}
\label{sec:coupling}
To explore whether the interactions in Eqs.~(\ref{eq:lag_P}), (\ref{eq:lag_pi}), or (\ref{eq:lag_S})
 can explain the rise of the $\pi^0$-$\gamma$ form factor 
in agreement with behavior seen by \babar, we first have to choose a 
description of this form factor in QCD.  For the purposes of our fits we take the form 
factor obtained by a calculation of the DA using QCD sum rules with nonlocal condensates by Bakulev, Mikhailov, and Stefanis (BMS)~\cite{Bakulev:2001pa,*Bakulev:2001pb}.  BMS obtained a range of allowable form factors by varying the cutoff of an integral of a perturbative spectral function by $\pm10\%$ in their sum rule.  We allow the BMS form factor that we use in our fits to vary in this range.

In the case of the pseudoscalar impostor we fit the data on ${\cal F}\left(Q^2\right)$ from \babar, CLEO, and CELLO using [see Eq.~(\ref{eq:PImPP_FF})]
\begin{align}
\left|{\cal F}\left(Q^2\right)\right|=\sqrt{\left|{\cal F}_{\rm BMS}\left(Q^2\right)\right|^2+\left|{\cal F}_{P}\left(Q^2\right)\right|^2}
\label{eq:fit_Im}
\end{align}
with ${\cal F}_{\rm BMS}$ the BMS form factor described above, which accounts for the QCD contribution, and ${\cal F}_P$ from Eq.~(\ref{FFP}) with $m_P=m_{\pi^0}$.  ${\cal F}_P$ depends very weakly on $m_P$ changing by less than 1\% for $m_{\pi^0}/2<m_P<2m_{\pi^0}$ in the relevant $Q^2$ range.  For the scalar impostor we do the same, replacing ${\cal F}_P$ with ${\cal F}_S$ from Eq.~(\ref{FFS}) with $m_S=m_{\pi^0}$.  We fit [Eq.~(\ref{eq:HCP_FF})]
\begin{align}
\left|{\cal F}\left(Q^2\right)\right|=\left|{\cal F}_{\rm BMS}\left(Q^2\right)+{\cal F}_{\rm HC\pi}\left(Q^2\right)\right|
\label{eq:fit_hcp}
\end{align}
for the hardcore pion, where ${\cal F}_{\rm HC\pi}$ is given by Eq.~(\ref{FFP}) with $g_\tau\to g_\tau^\pi$ and $m_P=m_{\pi^0}$.

The results of these fits, along with the range of BMS form factors used, are shown in Fig.~\ref{fig:FF}.
\begin{figure}
\rotatebox{270}{\resizebox{70mm}{!}{\includegraphics{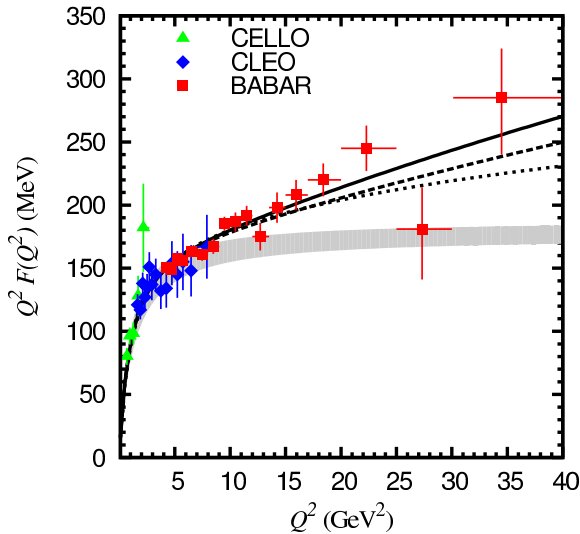}}}
\caption{The best-fit total form factors in the pseudoscalar impostor (solid), scalar impostor (dashed), and hardcore pion (dotted) cases when fit to the data from \babar~\cite{Aubert:2009mc}, CLEO~\cite{Gronberg:1997fj}, and CELLO~\cite{Behrend:1990sr} using the form factors of Eqs.~(\ref{eq:fit_Im}) and (\ref{eq:fit_hcp}).  The resultant couplings are $g_\tau=0.63$, $h_\tau=0.84$, and $g_\tau^\pi=0.18$, see Table~\ref{tab:couplings}.  The shaded region indicates the range of BMS form factors.  The scalar impostor coupling is limited by the $(g-2)_\tau$ constraint~\cite{Abdallah:2003xd}.}
\label{fig:FF}
\end{figure}
The range of BMS form factors corresponds to a deviation from the data of at least $3.5\sigma$.

In the pseudoscalar impostor case we find a coupling $g_\tau=0.63^{+0.17}_{-0.22}$.  The best-fit coupling deviates from the data at $0.8\sigma$ level and the total form factor using this coupling is shown by the solid line in Fig.~\ref{fig:FF}.

The fit in the scalar impostor case gives $h_\tau=0.97^{+0.26}_{-0.34}$.  Much of this region conflicts with the constraint from $(g-2)_\tau$; taking this into account we arrive at a best-fit coupling $h_\tau=0.84$ which fits the data at the $0.9\sigma$ level. The total form factor with this coupling is shown by the dashed line in Fig.~\ref{fig:FF}.

Fitting the data in the hardcore pion case results in $g_\tau^\pi=0.18^{+0.10}_{-0.12}$. The best-fit coupling here, which corresponds to a $1.6\sigma$ deviation from the data, is smaller since there is constructive interference with the BMS form factor which describes the data quite well at low $Q^2$.  In Fig.~\ref{fig:FF}, we show the total form factor using the best-fit coupling in this case with the dotted line .

The pseudoscalar impostor fits the data best of the three scenarios with the scalar impostor running into some tension with the limits from $(g-2)_\tau$ and the hardcore pion being constrained by the good agreement of the BMS form factor with the low $Q^2$ data.

The situation is not drastically different if we restrict our fit only to larger $Q^2$ data.  For $Q^2>8~{\rm GeV}^2$, the range of BMS form factors deviates by at least $4.9\sigma$ from the data.  Fits result in $g_\tau=0.70^{+0.17}_{-0.25}$, $h_\tau=1.08^{+0.26}_{-0.38}$, and $g_\tau^\pi=0.25^{+0.06}_{-0.14}$.  Again, $h_\tau=0.84$ is chosen as the best-fit coupling in the scalar impostor case due to $(g-2)_\tau$ constraint.  The best-fit couplings deviate from the data at the $1.1\sigma$, $1.5\sigma$, and $1.2\sigma$ levels for the pseudoscalar impostor, scalar impostor, and hardcore pion cases, respectively.

We summarize the best-fit couplings and the level of their agreement with the data in Table~\ref{tab:couplings}.
\begin{table*}
\begin{tabular}{c|ccc|ccc}
\hline
\hline
& \multicolumn{3}{c|}{$Q^2>0~{\rm GeV}^2$} & \multicolumn{3}{c}{$Q^2>8~{\rm GeV}^2$}\tabularnewline
\cline{2-7}
& coupling$\pm1\sigma$ & $\chi^2$/d.o.f. & agreement & coupling$\pm1\sigma$ & $\chi^2$/d.o.f. & agreement \tabularnewline
\hline
\hline
$P$ & $0.63^{+0.17}_{-0.22}$ & 35.5/35 & $0.8\sigma$ & $0.70^{+0.17}_{-0.25}$ & 10.8/9 & $1.1\sigma$ \tabularnewline
$S$ & $0.97^{+0.26}_{-0.34}\to 0.84$ & 37.8/35 & $0.9\sigma$ & $1.08^{+0.26}_{-0.38}\to 0.84$ & 13.4/9 & $1.5\sigma$ \tabularnewline
HCP & $0.18^{+0.10}_{-0.12}$ & 45.3/35 & $1.6\sigma$ & $0.25^{+0.06}_{-0.14}$ & 11.9/9 & $1.2\sigma$ \tabularnewline
\hline
BMS & & 72.7/36 & $3.6\sigma$ & & 49.3/10 & $5.1\sigma$ \tabularnewline
\hline
\hline
\end{tabular}
\caption{The $\pm1\sigma$ range of couplings ($g_\tau$, $h_\tau$, $g_\tau^\pi$), from Eqs.~(\ref{eq:lag_P}), (\ref{eq:lag_pi}), and (\ref{eq:lag_S}), in the pseudoscalar impostor ($P$), scalar impostor ($S$), and hardcore pion (HCP) cases when fit to the $\pi^0$-$\gamma$ form factor data in the $Q^2$ ranges indicated.  We also show the $\chi^2$/d.o.f. and the corresponding level of agreement with the data for the best-fit value in each case.  Note that the best-fit value of coupling in the scalar impostor case is limited to the largest fit value (0.84) compatible with the $(g-2)_\tau$ constraint~\cite{Abdallah:2003xd}--the $\chi^2$ reported here is with respect to this value.  Also shown is the level of agreement between the data and the BMS form factors~\cite{Bakulev:2001pa,*Bakulev:2001pb} for each $Q^2$ range.  The total $\pi^0$ transition form factors using the best-fit coupling in each case are shown in Fig.~\ref{fig:FF}.}
\label{tab:couplings}
\end{table*}
\end{subsection}


\begin{subsection}{Impostor Mass}
\label{sec:mass}
It is clear that in order for pseudoscalar or scalar impostors coupled to $\tau$ to offer an explanation of 
the experimentally seen rise in the $\pi^0$ form factor as in Sec.~\ref{sec:coupling}, they must promptly decay to $\gamma\gamma$.
Also, $m_\pi^0-m_{P,S}$ must be sufficiently small so that these new particles are reconstructed as 
neutral pions in the reactions $e^+e^-\to e^+e^-(P,S)$. This will create an apparent contribution 
to the cross section for $e^+e^-\to e^+e^-\pi^0$, which is used to extract the form factor.  
We would like to understand just how small $m_\pi^0-m_{P,S}$ is required to be.

Using the best-fit couplings in the impostor cases above, along with the expressions for the cross sections for $e^+e^-\to e^+e^-(P,S,\pi^0)$ as functions of the form factors for $\gamma\gamma^\ast\to(P,S,\pi^0)$, we can estimate the number of actual pions and impostors produced in a given $Q^2$ range.  For instance, using the \babar~experimental setup and the best-fit coupling for the pseudoscalar impostor, $g_{\tau}=0.63$, implies that the ratio of the numbers impostors to neutral pions produced is $N_P/N_{\pi^0}=5.4\%$ for $Q^2=4-4.5~{\rm GeV}^2$.  This ratio is quite insensitive to the mass of the pseudoscalar, varying by less than a factor of $10^{-2}$ as $m_P$ ranges from $m_{\pi^0}/2$ to $2m_{\pi^0}$.  We then fit the $M_{\gamma\gamma}$ spectrum obtained by \babar~in this $Q^2$ range (Fig.~6(a) of Ref.~\cite{Aubert:2009mc}), which is used to estimate the $\pi^0$ production cross section, with two Gaussians, one centered at $m_{\pi^0}$ and the other at a value of $m_P$ that we choose, on top of a quadratic background.  We fix the widths of both to $9~{\rm MeV}$, set by the experimental photon energy resolution.  Their heights are then related by $N_P/N_{\pi^0}$ ($5.4\%$ here).  We repeat these fits while scanning $m_P$ over a broad range around $m_{\pi^0}$ and record the change in $\chi^2$ from the value obtained when $m_P=m_{\pi^0}$.  The allowed range of $m_P$ will be given by masses that do not lead to unacceptably large $\chi^2$ when fitting the measured $M_{\gamma\gamma}$ spectrum.  We then perform the same fits of the \babar~data for $Q^2=11-12~{\rm GeV}^2$ and $Q^2=20-25~{\rm GeV}^2$ where $N_P/N_{\pi^0}=24\%,~62\%$ respectively using the spectra from Figs.~6(b) and (c) of Ref.~\cite{Aubert:2009mc}.  The ratios $N_P/N_{\pi^0}$ are similarly insensitive to the particular $m_P$ as for $Q^2=4-4.5~{\rm GeV}^2$. 

The results of these fits are shown in Fig.~\ref{fig:massP}.  The $Q^2=20-25~{\rm GeV}^2$ data is most constraining on the allowed range of $m_P$ although there are fewer events than in the lower $Q^2$ ranges, due to the increased ratio of pseudoscalar impostors produced.  There is a roughly $\pm 5~{\rm MeV}$ range near $m_{\pi^0}$ where the $\chi^2$ is not increased by more than a few in each $Q^2$ range.  We note that there are $51-56$ degrees of freedom in each fit.
\begin{figure}
\rotatebox{270}{\resizebox{70mm}{!}{\includegraphics{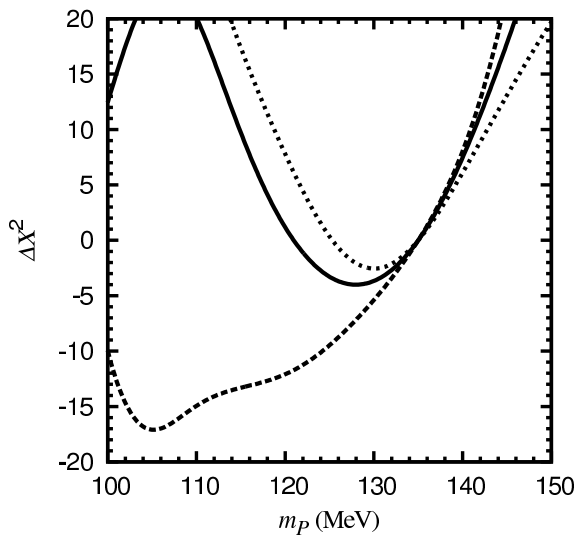}}}
\caption{Constraining the pseudoscalar impostor's mass: the change in $\chi^2$ fitting the $M_{\gamma\gamma}$ spectra of Fig.~6 of Ref.~\cite{Aubert:2009mc} to Gaussians centered at $m_{\pi^0}$ and $m_P$ with a quadratic background, relative to the $\chi^2$ for $m_P=m_{\pi^0}$.  The solid curve is for $Q^2=4-4.5~{\rm GeV}^2$, the dashed for $Q^2=11-12~{\rm GeV}^2$, and the dotted for $Q^2=20-25~{\rm GeV}^2$.  The heights of the two Gaussians used in the fits are related by the values of $N_P/N_{\pi^0}$ in each $Q^2$ bin (5.4\%, 24\%, and 62\%).  There are $51-56$ degrees of freedom in each fit.}
\label{fig:massP}
\end{figure}

We repeat this procedure for the scalar with its best-fit coupling of $h_\tau=0.84$ and show the changes in $\chi^2$ in Fig.~\ref{fig:massS}.  The ratios of impostors to pions we use is $N_S/N_{\pi^0}=4.2\%,~18\%,~46\%$ for $Q^2=4-4.5,~11-12,~20-25~{\rm GeV}^2$ respectively.  There are fewer impostors produced in this case compared to the pseudoscalar one, as can be seen in the smaller form factor in Fig.~\ref{fig:FF}, and so the constraints on $m_S$ are less tight with $\chi^2$ not increasing by more than a few over a range of about $\pm 8~{\rm MeV}$ around $m_{\pi^0}$.  
\begin{figure}
\rotatebox{270}{\resizebox{70mm}{!}{\includegraphics{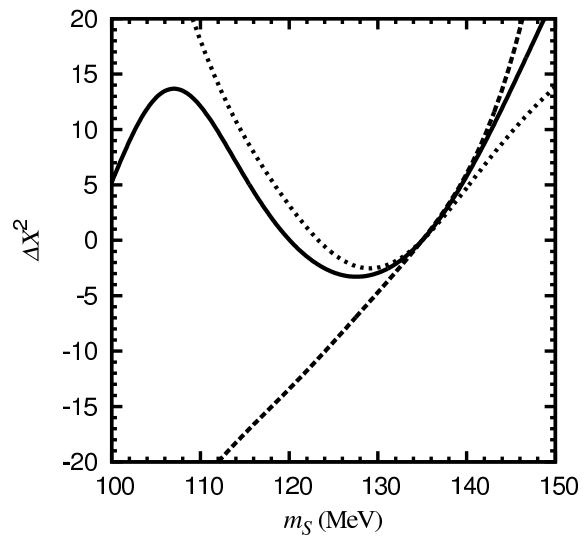}}}
\caption{Similar to Fig.~\ref{fig:massP} but in the scalar impostor case.  $N_S/N_{\pi^0}=4.2\%,~18\%,~46\%$ for $Q^2=4-4.5,~11-12,~20-25~{\rm GeV}^2$.}
\label{fig:massS}
\end{figure}
\end{subsection}

\end{section}


\begin{section}{Further Constraints and Tests}
\label{sec:constraints}
As seen in Sec.~\ref{sec:fits}, explaining the deviation between the measured $\pi^0$-$\gamma$ form factor and most expectations in QCD with a coupling of the $\tau$ to $\pi^0$ or to a new pseudoscalar or scalar with a mass within about $5~{\rm MeV}$ of $m_{\pi^0}$ requires relatively large couplings, of the order $0.1-1$.  We would like to see if such interactions are constrained by different measurements and, if not, how to probe them.


\begin{subsection}{Anomalous Magnetic Moments}
We have seen in Secs.~\ref{sec:g-2} and \ref{sec:fits} that the current experimental limit on $(g-2)_\tau$ is not very constraining in the case of a pseudoscalar impostor or a coupling between $\tau$ and $\pi^0$.  However, it does limit the scalar impostor coupling to $\tau$ to $h_\tau=0.84$, causing it to not fit the rise in the $\pi^0$-$\gamma$ form factor as well as the pseudoscalar. 

In addition to the one-loop contribution to $(g-2)_\tau$, interactions like those in Eqs.~(\ref{eq:lag_P}), (\ref{eq:lag_pi}), and (\ref{eq:lag_S}) would contribute to $(g-2)_{e,\mu}$ at the three-loop level suppressed by at least a loop factor $\sim 1/16\pi^2$ times a coupling squared relative to the leading two-loop contribution involving a $\tau$ loop.

The leading $\tau$ contribution to $a_\mu$ is roughly $4.2\times10^{-10}$,  comparable to the current experimental and theoretical uncertainties~\cite{Jegerlehner:2009ry}.  Therefore, the three-loop contribution is safely below the uncertainty and we conclude that ${\cal O}(1)$ couplings of the $\tau$ to a hardcore pion or impostors are not ruled out by corrections to $(g-2)_{\mu}$.

In the case of the $a_e$, the leading $\tau$ contribution is down by a factor $\sim m_e^2/m_\mu^2$ relative to the muon case.  Thus, the three-loop contribution to $a_e$ from an impostor or hardcore pion coupling to $\tau$ is well below the upper limit of $15\times10^{-9}$ from other precision measurements of $\alpha$ remaining consistent with $a_e$~\cite{Hanneke:2008tm,*Clade:2006zz,*Gerginov:2006zz}.

In passing, we note that a $\pi^0$ transition form factor that rises relative to $Q^{-2}$ at the level implied by the \babar\ data does not contribute to a meaningful enhancement in the hadronic light-by-light contribution to $(g-2)_{\mu}$~\cite{Nyffeler:2009uw}.
\end{subsection}


\begin{subsection}{$\pi^0$ Width}
\label{sec:pi0width}
Currently, the most precise measurement of the $\pi^0\to\gamma\gamma$ rate, to $2.8\%$, has been made by the PrimEx experiment~\cite{Larin:2010kq}, using the measurement of the $\pi^0$ production cross section through the Primakoff process $\gamma\, {\rm Nuc.}\to\pi^0 {\rm Nuc.}$ which can be related to $\Gamma\left(\pi^0\to\gamma\gamma\right)$.  A hardcore pion coupling to $\tau$ or impostors with masses close to that of the $\pi^0$ could lead to an apparent deviation between this lifetime measurement from theoretical calculations.

Using the limit for the QCD $\pi^0$ form factor at $Q^2=0$ from the chiral anomaly, and the expression for the hardcore pion form factor in Eq.~\ref{eq:HCP_FF}, we find a fractional change in the rate for $\pi^0\to\gamma\gamma$ of
\begin{align}
\frac{\delta\Gamma_{\pi^0\to\gamma\gamma}}{\Gamma_{\pi^0\to\gamma\gamma}}&\simeq g_\tau^\pi\frac{\sqrt{2} f_\pi}{m_\tau}\left(1+\frac{g_\tau^\pi f_\pi}{2\sqrt{2}m_\tau}\right)\simeq 0.1 g_\tau^\pi.
\end{align}
Using the best-fit value $g_\tau^\pi=0.18$ gives a change of about $1.8\%$.  If we use the value of $g_\tau^\pi$ obtained from fitting only the $Q^2>8~{rm GeV}^2$ data it becomes $2.5\%$.  These shifts are smaller than the experimental uncertainty as well as the uncertainty of theoretical calculations, 
and thus these values of $g_\tau^\pi$ are not ruled out.

In the case of a pseudoscalar impostor, the apparent shift in the lifetime is simply
\begin{align}
\frac{\delta\Gamma_{\pi^0\to\gamma\gamma}}{\Gamma_{\pi^0\to\gamma\gamma}}&=\frac{\Gamma_{P\to\gamma\gamma}}{\Gamma_{\pi^0\to\gamma\gamma}}\simeq\frac{1}{\Gamma_{\pi^0\to\gamma\gamma}}\frac{g_\tau^2\alpha^2m_{\pi^0}^3}{64\pi^3m_\tau^2}
\\
&\simeq 2.6\times 10^{-3}g_\tau^2.
\end{align}
In the case of a scalar impostor this is instead
\begin{align}
\frac{\delta\Gamma_{\pi^0\to\gamma\gamma}}{\Gamma_{\pi^0\to\gamma\gamma}}&\simeq\frac{1}{\Gamma_{\pi^0\to\gamma\gamma}}\frac{h_\tau^2\alpha^2m_{\pi^0}^3}{144\pi^3m_\tau^2}\simeq 1.2\times 10^{-3}h_\tau^2.
\end{align}
Using the best-fit values for these couplings leads to shifts that are below the experimental uncertainty and we conclude that they are also allowed.

Lastly, following~\cite{Dorokhov:2010zz}, we do not expect a large contribution to the rate for $\pi^0\to e^+e^-$ in any of the scenarios we have considered.
\end{subsection}


\begin{subsection}{$Z$ Decays}
A hardcore pion or impostor coupling to $\tau$ would also lead to (or mimic) $Z\to\gamma\pi^0$ decays through a $\tau$ loop.  However, because of the suppressed vector coupling of the $Z$ to $\tau$, these processes would contribute to this decay mode with a branching less that $10^{-8}$ for couplings $g_\tau$, $h_\tau$, $g_\tau^\pi$ equal to unity, well below the current limit of $5.2\times10^{-5}$~\cite{Acciarri:1995gy}.

An associated emission of impostor particles or hardcore pions can be also constrained by the 
non-observation of $Z\to \tau^+\tau^-\gamma$ events by OPAL \cite{Acton:1991dq} with  
$\gamma$ well-separated from $\tau$ leptons. Since at high energy the
signal from pion-like particles will be indistinguishable from $\gamma$, 
we calculate the branching ratio of the $Z$ to $\tau$ pairs plus an extra particle,  
limit its energy to be $E_{P,S,\pi^0} > 1~{\rm GeV}$, require that $\cos\theta<0.95$ where $\theta$ is the smaller of the angles between $(P,S,\pi^0)$ and $\tau^+$ or $\tau^-$, the 
constraint ${\cal B}(Z\to \tau^+\tau^-(P,S,\pi^0))$ to be less than the experimental 
bound on the branching to $\tau^+\tau^-\gamma$ of $7.3\times 10^{-4}$.  The resulting sensitivity to
coupling constants is
\begin{equation}
\left|g_\tau\right|,\left|g_\tau^\pi\right|\lsim1.0;~~~\left|h_\tau\right|\lsim1.1.
\label{eq:sensitivity}
\end{equation}
In order to convert this to strict constraints we need to choose energy $E_{P,S,\pi^0}$, where 
$\pi^0$ and $\gamma$ signals cannot be differentiated. Choosing this energy to be $\sim 10$ GeV 
further relaxes the limits in Eq.~(\ref{eq:sensitivity}) by a factor of a few. We conclude that such 
 limits are not quite sensitive enough to test the couplings required for a pion impostor or hardcore pion to explain the form factor measurement of \babar.

\end{subsection}


\begin{subsection}{Searching for an Impostor}
\begin{subsubsection}{$\tau$ Decays}
One obvious place to look for a hardcore pion or pion impostors that interact with the $\tau$ as in Eqs.~(\ref{eq:lag_P}), (\ref{eq:lag_pi}), and (\ref{eq:lag_S}) is in $\tau$ decays.  It is well known that hadronic decays of $\tau$ can be used a way of extracting the hadronic vacuum polarization,
and that $\tau$-based extraction gives larger result than the direct $e^+e^-\to {\rm hadrons}$ by about a few percent \cite{Davier:2009ag}. 
This mild tension provides additional motivation to search for models with additional ``hadron-like" particles coupled to $\tau$-leptons. 

Our models will lead to unusual 
decay channels such as $\tau^\pm\to \pi^0\ell^\pm\bar\nu\nu$ or 
$\tau^\pm\to P\ell^\pm\bar\nu\nu$ and $\tau^\pm\to S\ell^\pm\bar\nu\nu$, the latter two mimicking the first. Here
 $\ell$ denotes final state electrons and muons.  The branching ratios are estimated using CalcHEP~\cite{Pukhov:2004ca,*Pukhov:1999gg} and found to be
\begin{align}
{\cal B}\left(\tau^\pm\to \pi^0 e^\pm\bar\nu\nu\right)&\simeq 3.5\times 10^{-5}\left(g_\tau^\pi\right)^2,
\\
{\cal B}\left(\tau^\pm\to P e^\pm\bar\nu\nu\right)&\simeq 3.5\times 10^{-5}g_\tau^2,
\\
{\cal B}\left(\tau^\pm\to S e^\pm\bar\nu\nu\right)&\simeq 1.7\times 10^{-3}h_\tau^2.
\end{align}
The best-fit values of the couplings lead to
\begin{align}
{\cal B}\left(\tau^\pm\to \pi^0 e^\pm\bar\nu\nu\right)&\simeq 1.1\times 10^{-6},
\\
{\cal B}\left(\tau^\pm\to P e^\pm\bar\nu\nu\right)&\simeq 1.4\times 10^{-5},
\\
{\cal B}\left(\tau^\pm\to S e^\pm\bar\nu\nu\right)&\simeq 1.2\times 10^{-3}.
\end{align}

The sum of all tau branching fractions,
 $0.9984\pm0.0010$~\cite{Asner:2010qj},
 although consistent with unity does allow for
 ${\cal B}\left(\tau^\pm\to (P,S,\pi^0) \ell^\pm\bar\nu\nu\right)<3.6\times 10^{-3}$ at 95\% confidence level.
Such modes would not have been detected in conventional $\tau$ branching fraction 
measurements and the related tau hadronic modes that include $(P, S, \pi^0)$ would not have been detected as they would be included
unknowingly in conventional hadronic branching fraction studies. Only a dedicated search for  
 $\tau^\pm\to (P,S,\pi^0) \ell^\pm\bar\nu\nu$ would reveal its presence in $\tau$ lepton decay and
to the best of our knowledge no such experiment has been performed. For example,
due to the neutrinos in the final state, these decays are not ruled out by the limits on the branching ratios of lepton flavor-violating (LFV) decays $\tau^\pm\to\ell^\pm\pi^0$ at the $10^{-8}-10^{-7}$ level~\cite{Aubert:2006cz,*Miyazaki:2007jp} since these searches require that the invariant mass of the $\ell\pi^0$ system be consistent with $m_\tau$.  A loose requirement that the invariant mass of the lepton-impostor pair be greater than $1.5~{\rm GeV}$ suppresses these branching ratios by a factor of about $4\times10^{-4}$ in the pseudoscalar case and $5\times10^{-5}$ in the scalar case, bringing them below the experimental limits on LFV decays.

A difficulty in searching for these decays, especially for final states involving muons, is possible contamination from normal hadronic $\tau^\pm\to\rho^\pm\nu$ decays where $\rho^\pm\to\pi^\pm\pi^0$ if the charged pion fakes a lepton.
\end{subsubsection}

\begin{subsubsection}{Associated Production}
One could also search for $\tau$ pair production in association with an impostor in $e^+e^-$ collisions.  At $\sqrt s=10.58~{\rm GeV}$ the cross sections for this are
\begin{align}
\sigma\left(e^+ e^- \to \tau^+\tau^- \pi^0\right)&\simeq 9.8\ \left(g_\tau^\pi\right)^2~{\rm pb},
\\
\sigma\left(e^+ e^- \to \tau^+\tau^- P\right)&\simeq 9.8\ g_\tau^2~{\rm pb},
\\
\sigma\left(e^+ e^- \to \tau^+\tau^- S\right)&\simeq 110\ h_\tau^2~{\rm pb}.
\end{align}
While these cross sections are large, it is difficult to separate the impostors or hardcore pion from the neutral pions produced in hadronic tau decays.   As far as we know, such searches have not been performed.  The kinematics of these final states can offer some advantages in searching for impostors or a hardcore pion when compared with searches in $\tau$ decays.

A large background to $\tau^+\tau^- (\pi^0,P,S)$ with both $\tau$'s decaying to leptons comes from $\tau^+\tau^-$ production where one of the $\tau$'s decays leptonically and the other to $\pi^\pm\pi^0$ and a neutrino with the $\pi^\pm$ being misidentified as a lepton.  To reduce such backgrounds, one can require that
\begin{align}
&E_{\rm large}+E_{\pi^0,P,S}>\frac{E_{\rm CM}}{2},
\label{eq:largecut}
\\
&E_{\rm small}+E_{\pi^0,P,S}>\frac{E_{\rm CM}}{2},
\label{eq:smallcut}
\end{align}
where $E_{\rm large}$ $(E_{\rm small})$ is the larger (smaller) of the charged lepton or pion energies.  This cut ensures that the leptons (or charged pion faking a lepton) and the impostor (or hardcore pion) cannot have all originated from $\tau$ decays following $\tau^+ \tau^-$ production.  To illustrate this, we generate $\tau^+ \tau^- (\pi^0,P,S)$ events with CalcHEP at $\sqrt s=10.58~{\rm GeV}$ with each $\tau$ decaying to electrons, $\tau^\pm\to e^\pm\bar\nu\nu$, to minimize the pion misidentification rate. 
In order to demonstrate the effect of background  from $e^+ e^- \to \tau^+\tau^-$ events,
we also generate $\tau^+ \tau^-$ events using PYTHIA 6.4~\cite{Sjostrand:2006za} with one $\tau$ decaying to an electron, $\tau^\pm\to e^\pm\bar\nu\nu$, and the other to $\pi^\pm\pi^0$ through a $\rho$, $\tau^\pm\to \rho^\pm\nu\to \pi^\pm\pi^0\nu$ and assign an electron misidentification rate of 2\% due to charged pions.  We require that $E_{\rm large}+E_{P,S}>E_{\rm CM}/2$ and plot the resulting cross sections as functions of $E_{\rm small}+E_{P,S}$ in Fig.~\ref{fig:events}.  About $5\%$ of $\tau^+\tau^- P$ events satisfy both Eqs.~(\ref{eq:largecut}) and (\ref{eq:smallcut}) as well as around $1\%$ of $\tau^+\tau^- S$ events while none of the events from the $\tau$ pair production background satisfy both of these cuts.
\begin{figure}
\rotatebox{270}{\resizebox{70mm}{!}{\includegraphics{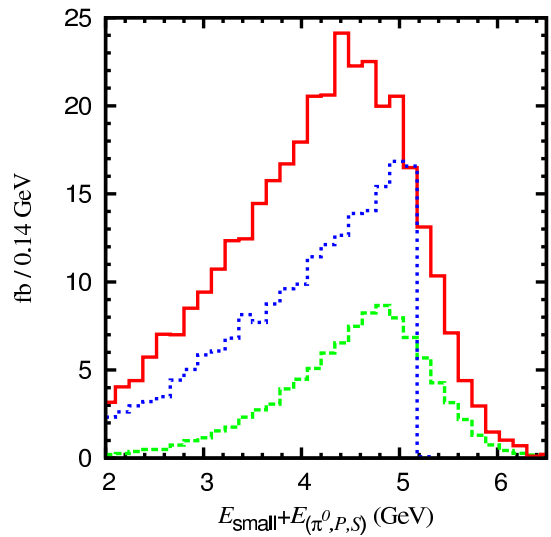}}}
\caption{The cross sections for $\tau^+\tau^-S$ (solid, red) with $h_\tau=1$ and $\tau^+\tau^-(\pi^0,P)$ (dashed, green) with $g_\tau^\pi=1$ and $g_\tau=1$, with both $\tau$'s decaying to electrons, and from $\tau^+\tau^-\to e^\pm \pi^\mp\pi^0\nu\bar\nu\nu$ (dotted, blue) with $\pi^\pm$ misidentified as $e^\pm$ 2\% of the time, shown as functions of $E_{\rm small}+E_{(\pi^0,P,S)}$.  The cut $E_{\rm large}+E_{(\pi^0,P,S)}>E_{\rm CM}/2=5.29~{\rm GeV}$ has been applied.  $E_{\rm small}$ is the smaller value of the charged lepton (or pion) energies and $E_{\rm large}$ the larger.  $E_{(\pi^0,P,S)}$ is the impostor or $\pi^0$ energy.  The $\tau^+\tau^-$ background has a cutoff at $E_{\rm small}+E_{(\pi^0,P,S)}=E_{\rm CM}/2=5.29~{\rm GeV}$ while the $\tau^+\tau^-(\pi^0,P,S)$ signals extend beyond that value.}
\label{fig:events}
\end{figure}
After these cuts, the effective cross sections for impostor or hardcore pion production become
\begin{align}
\left.\sigma\left(\tau^+\tau^- \pi^0\to e^+e^- 4\nu \pi^0\right)\right|_{\rm cuts}&\simeq 15\ \left(g_\tau^\pi\right)^2~{\rm fb},
\label{eq:hcp_signal}
\\
\left.\sigma\left(\tau^+\tau^- P\to e^+e^- 4\nu P\right)\right|_{\rm cuts}&\simeq 15\ g_\tau^2~{\rm fb},
\label{eq:p_signal}
\\
\left.\sigma\left(\tau^+\tau^- S\to e^+e^- 4\nu S\right)\right|_{\rm cuts}&\simeq 35\ h_\tau^2~{\rm fb}.
\label{eq:s_signal}
\end{align}
Using the best-fit couplings from the $\pi^0$ transition form factor fits gives effective cross sections after these cuts in the range $0.5-25~{\rm fb}$,
which would of course translate to $O(10^3)$ events in the full \babar\ dataset.   Probing cross sections at this level will decisively test whether new (pseudo)scalars coupled to $\tau$ are responsible for the observed rise in the $\pi^0$ form factor.

There is also a potential background from $e^+e^-\to\gamma\pi^+\pi^-\pi^0$ with the photon traveling along the beam direction, going undetected.  However, this can be eliminated by requiring missing transverse momentum.  This also reduces many of the light quark-induced backgrounds generally.

Charm pair production backgrounds, {\em e.g.}, $D^+D^-$ production, are reduced by the cuts in Eqs.~(\ref{eq:largecut}) and (\ref{eq:smallcut}), just as the $\tau^+\tau^-$ background is.  A potentially irreducible background comes from $\pi^0 D^+D^-$ and $\pi^\pm D^\mp D^0$ production with the $D$ mesons decaying to leptons and neutrinos, neutral pions, or to misidentified charged pions.  We estimate the production cross section for these final states to be of the order $10~{\rm pb}$.  The branching ratios to these final states are quite suppressed, especially to $e^\pm$, which is another motivation for choosing $\tau$ decays to electrons in this search.  There is further suppression from the misidentification rate as well from the cuts in Eqs.~(\ref{eq:largecut}) and (\ref{eq:smallcut}), causing these backgrounds to contribute at a level well below the relevant values of the signal cross section in Eqs.~(\ref{eq:hcp_signal})-(\ref{eq:s_signal}).

\end{subsubsection}
\begin{subsubsection}{Effect on Hadronic $\tau$ Decays}
The coupling of an impostor to $\tau$ could lead to an apparent increase in the rate for $\tau^\pm\to\pi^0\pi^\pm\nu$, an extremely important channel in estimating the hadronic spectral function used to calculate $(g-2)_\mu$.  This could happen if a $\tau$ emits an impostor in the 
three-body decay, $\tau^{\pm} \to (P,S)\pi^\pm\nu$.  It could also occur if a $\tau$ pair is produced in association with an impostor and one of the $\tau$'s decays to $\pi^\pm\nu$.  Taking both effects into account, we estimate a shift in this rate of
\begin{align}
\frac{\delta\Gamma\left(\tau^\pm\to\pi^0\pi^\pm\nu\right)}{\Gamma\left(\tau^\pm\to\pi^0\pi^\pm\nu\right)}&\simeq 0.26\%\ g_\tau^2
\\
\frac{\delta\Gamma\left(\tau^\pm\to\pi^0\pi^\pm\nu\right)}{\Gamma\left(\tau^\pm\to\pi^0\pi^\pm\nu\right)}&\simeq 3.2\%\ h_\tau^2.
\end{align}
The shift in the hardcore pion case is more difficult to estimate, due to the interference with the standard decay amplitude.

It is intriguing that this shift is not far from the level necessary to explain the longstanding discrepancy between the hadronic spectral functions extracted from $\tau$ decay and $e^+e^-$ data, see, {\em e.g.}, \cite{Davier:2009ag}.  This discrepancy is troubling since the hadronic spectral function provides the largest theoretical uncertainty in the estimate of $(g-2)_\mu$.  We leave a more complete analysis of this effect and the potential for it to alleviate the tension between the different determinations of $(g-2)_\mu$ for future work, but mention that such a solution appears possible.

In addition, an increase in the di-pion branching of the $\tau$ could change the measurement of the $V_{us}$ entry in the Cabibbo-Kobayashi-Maskawa (CKM) matrix using the ratio of strange to non-strange $\tau$ decays.  Currently, the value of $V_{us}$ extracted in this way is $3.3\sigma$ to $3.6\sigma$ smaller than the value predicted by the unitarity of the CKM matrix~\cite{Asner:2010qj,Banerjee:2011az}.  This sign of the change to $V_{us}$ looks correct in the models presented but more detailed study is required for quantitative conclusions.
\end{subsubsection}
\end{subsection}


\end{section}


\begin{section}{Conclusions}
\label{sec:conclusions}

The \babar\ measurement of the pion transition form factor did not agree with any of the QCD-based predictions that existed prior 
to the experimental publication. Experts' opinions are split whether the observed behavior can be accommodated within QCD even {\em a posteriori}, by, {\em e.g.}, drastically modifying the pion distribution amplitude. 
Alternatively, quark models could provide a reasonable fit to data, but would 
require that the
quark mass scale is on the order of 130 {\rm MeV}, which is well above the current quark mass values. We have shown that this quark mass 
scale may be emergent as a nonperturbative correction due to the vacuum condensate-induced mixing between the massless quark and pion 
propagators, $m^{\rm eff}_q \sim -\langle \bar qq\rangle/(16F_\pi^2)$. However, in order to explain the
experimental rise of $Q^2{\cal F}(Q^2)$, one has to take the chiral quark model well outside its usual range of validity. 
Both explanations of the \babar\ asymptotics of the pion form factor--QCD with a flattened distribution amplitude or a massive quark model--imply that the $\pi^0$ field is more elementary than originally thought.

The main focus of this paper is the new physics explanation of the form factor behavior, which is perhaps the least likely option 
among different logical possibilities outlined in the introduction. 
Nevertheless, we were able to construct some models of new physics 
that can be consistent with the observed form factor behavior. 
These models operate with new scalar 
or pseudoscalar particles coupled to $\tau$ leptons, affecting ${\cal F}_{\pi^0}$ at the one-loop level. 
The choice of coupling to the $\tau$ sector is motivated by necessity: {\em e.g.}, similar models with light particles coupled to other leptons or to
charm or bottom quarks are too constrained to play any role in the pion form factor. 
New particles can either imitate $\pi^0$, hence the name impostors, or lead to the enhanced coupling 
of $\pi^0$ to $\tau$'s, thus supplying pions with a ``hard core." 
Fitting experimental data, we were able to show that the range of couplings for such new states 
is consistent with existing constraints, while the mass of the impostors must be within a few MeV of $m_\pi$. 
This might look like fine-tuning, except that similar accidents happen with SM particles (such as, {\em e.g.}, the 5\% 
coincidence of $\tau$ lepton and $D$ meson masses).  The closeness of $m_{\rm impostor}$
to the pion mass would naturally explain why there is no associated enhancement of the form factor for $\eta$ and $\eta^\prime$.

A relatively large coupling of new states to $\tau$ leptons means that they can be searched for in a sizable $\tau$ dataset. 
We have shown that both $\tau$ decays, and especially associated production $e^+e^-\to \tau^+\tau^- (P,S,\pi^0)$
can be used to search for new ``pion-like" physics. Choosing the optimal couplings from the fits to the pion form factor, we find that 
$B$-factory samples should have on the order of $O(10^3)$ events in the associated production channels with distinct kinematics 
that will facilitate their search. Such a large number of expected events gives hope that future experimental analyses can be 
conclusive: they either discover such new states or close perhaps the only new physics explanation of the unusual 
behavior of the pion transition form factor. 

Lastly, we note that interactions like those that we considered could provide an explanation of the tension between the hadronic vacuum polarization  extracted from $\tau$ and $e^+e^-$ data, which have important implications 
for the muon $g-2$. Furthermore, there could be implications for the extraction of $V_{us}$ from $\tau$ decays.  Because of the importance of these subjects and detailed experimental data in 
the di-pion channels, further 
studies of hadronic $\tau$ decays within these models might be warranted. 
\end{section}

\begin{acknowledgements}
First and foremost, the authors would like to 
acknowledge important contributions from Brian Batell, who participated on the initial 
stages of this project. DM and MP would like to thank S.~Brodsky, G.~King, A.~Ritz, and X.-G.~Wu for 
useful conversations. 
This work was supported in part by NSERC, Canada, and research at the Perimeter Institute
is supported in part by the Government of Canada 
through NSERC and by the Province of Ontario through MEDT.
\end{acknowledgements}

\bibliography{ref}

\end{document}